\documentclass[sigconf]{acmart}

\usepackage{amsmath,amsfonts}
\usepackage{algorithmic}
\usepackage{graphicx}
\usepackage{textcomp}
\usepackage{xcolor}
\usepackage[utf8]{inputenc}
\usepackage{censor}
\usepackage[draft]{minted}
\usepackage{hyperref}
\usepackage{enumitem}
\usepackage{draftwatermark}

\makeatletter
\def\PYG@reset{\let\PYG@it=\relax \let\PYG@bf=\relax%
    \let\PYG@ul=\relax \let\PYG@tc=\relax%
    \let\PYG@bc=\relax \let\PYG@ff=\relax}
\def\PYG@tok#1{\csname PYG@tok@#1\endcsname}
\def\PYG@toks#1+{\ifx\relax#1\empty\else%
    \PYG@tok{#1}\expandafter\PYG@toks\fi}
\def\PYG@do#1{\PYG@bc{\PYG@tc{\PYG@ul{%
    \PYG@it{\PYG@bf{\PYG@ff{#1}}}}}}}
\def\PYG#1#2{\PYG@reset\PYG@toks#1+\relax+\PYG@do{#2}}

\expandafter\def\csname PYG@tok@gd\endcsname{\def\PYG@tc##1{\textcolor[rgb]{0.63,0.00,0.00}{##1}}}
\expandafter\def\csname PYG@tok@gu\endcsname{\let\PYG@bf=\textbf\def\PYG@tc##1{\textcolor[rgb]{0.50,0.00,0.50}{##1}}}
\expandafter\def\csname PYG@tok@gt\endcsname{\def\PYG@tc##1{\textcolor[rgb]{0.00,0.27,0.87}{##1}}}
\expandafter\def\csname PYG@tok@gs\endcsname{\let\PYG@bf=\textbf}
\expandafter\def\csname PYG@tok@gr\endcsname{\def\PYG@tc##1{\textcolor[rgb]{1.00,0.00,0.00}{##1}}}
\expandafter\def\csname PYG@tok@cm\endcsname{\let\PYG@it=\textit\def\PYG@tc##1{\textcolor[rgb]{0.25,0.50,0.50}{##1}}}
\expandafter\def\csname PYG@tok@vg\endcsname{\def\PYG@tc##1{\textcolor[rgb]{0.10,0.09,0.49}{##1}}}
\expandafter\def\csname PYG@tok@vi\endcsname{\def\PYG@tc##1{\textcolor[rgb]{0.10,0.09,0.49}{##1}}}
\expandafter\def\csname PYG@tok@vm\endcsname{\def\PYG@tc##1{\textcolor[rgb]{0.10,0.09,0.49}{##1}}}
\expandafter\def\csname PYG@tok@mh\endcsname{\def\PYG@tc##1{\textcolor[rgb]{0.40,0.40,0.40}{##1}}}
\expandafter\def\csname PYG@tok@cs\endcsname{\let\PYG@it=\textit\def\PYG@tc##1{\textcolor[rgb]{0.25,0.50,0.50}{##1}}}
\expandafter\def\csname PYG@tok@ge\endcsname{\let\PYG@it=\textit}
\expandafter\def\csname PYG@tok@vc\endcsname{\def\PYG@tc##1{\textcolor[rgb]{0.10,0.09,0.49}{##1}}}
\expandafter\def\csname PYG@tok@il\endcsname{\def\PYG@tc##1{\textcolor[rgb]{0.40,0.40,0.40}{##1}}}
\expandafter\def\csname PYG@tok@go\endcsname{\def\PYG@tc##1{\textcolor[rgb]{0.53,0.53,0.53}{##1}}}
\expandafter\def\csname PYG@tok@cp\endcsname{\def\PYG@tc##1{\textcolor[rgb]{0.74,0.48,0.00}{##1}}}
\expandafter\def\csname PYG@tok@gi\endcsname{\def\PYG@tc##1{\textcolor[rgb]{0.00,0.63,0.00}{##1}}}
\expandafter\def\csname PYG@tok@gh\endcsname{\let\PYG@bf=\textbf\def\PYG@tc##1{\textcolor[rgb]{0.00,0.00,0.50}{##1}}}
\expandafter\def\csname PYG@tok@ni\endcsname{\let\PYG@bf=\textbf\def\PYG@tc##1{\textcolor[rgb]{0.60,0.60,0.60}{##1}}}
\expandafter\def\csname PYG@tok@nl\endcsname{\def\PYG@tc##1{\textcolor[rgb]{0.63,0.63,0.00}{##1}}}
\expandafter\def\csname PYG@tok@nn\endcsname{\let\PYG@bf=\textbf\def\PYG@tc##1{\textcolor[rgb]{0.00,0.00,1.00}{##1}}}
\expandafter\def\csname PYG@tok@no\endcsname{\def\PYG@tc##1{\textcolor[rgb]{0.53,0.00,0.00}{##1}}}
\expandafter\def\csname PYG@tok@na\endcsname{\def\PYG@tc##1{\textcolor[rgb]{0.49,0.56,0.16}{##1}}}
\expandafter\def\csname PYG@tok@nb\endcsname{\def\PYG@tc##1{\textcolor[rgb]{0.00,0.50,0.00}{##1}}}
\expandafter\def\csname PYG@tok@nc\endcsname{\let\PYG@bf=\textbf\def\PYG@tc##1{\textcolor[rgb]{0.00,0.00,1.00}{##1}}}
\expandafter\def\csname PYG@tok@nd\endcsname{\def\PYG@tc##1{\textcolor[rgb]{0.67,0.13,1.00}{##1}}}
\expandafter\def\csname PYG@tok@ne\endcsname{\let\PYG@bf=\textbf\def\PYG@tc##1{\textcolor[rgb]{0.82,0.25,0.23}{##1}}}
\expandafter\def\csname PYG@tok@nf\endcsname{\def\PYG@tc##1{\textcolor[rgb]{0.00,0.00,1.00}{##1}}}
\expandafter\def\csname PYG@tok@si\endcsname{\let\PYG@bf=\textbf\def\PYG@tc##1{\textcolor[rgb]{0.73,0.40,0.53}{##1}}}
\expandafter\def\csname PYG@tok@s2\endcsname{\def\PYG@tc##1{\textcolor[rgb]{0.73,0.13,0.13}{##1}}}
\expandafter\def\csname PYG@tok@nt\endcsname{\let\PYG@bf=\textbf\def\PYG@tc##1{\textcolor[rgb]{0.00,0.50,0.00}{##1}}}
\expandafter\def\csname PYG@tok@nv\endcsname{\def\PYG@tc##1{\textcolor[rgb]{0.10,0.09,0.49}{##1}}}
\expandafter\def\csname PYG@tok@s1\endcsname{\def\PYG@tc##1{\textcolor[rgb]{0.73,0.13,0.13}{##1}}}
\expandafter\def\csname PYG@tok@dl\endcsname{\def\PYG@tc##1{\textcolor[rgb]{0.73,0.13,0.13}{##1}}}
\expandafter\def\csname PYG@tok@ch\endcsname{\let\PYG@it=\textit\def\PYG@tc##1{\textcolor[rgb]{0.25,0.50,0.50}{##1}}}
\expandafter\def\csname PYG@tok@m\endcsname{\def\PYG@tc##1{\textcolor[rgb]{0.40,0.40,0.40}{##1}}}
\expandafter\def\csname PYG@tok@gp\endcsname{\let\PYG@bf=\textbf\def\PYG@tc##1{\textcolor[rgb]{0.00,0.00,0.50}{##1}}}
\expandafter\def\csname PYG@tok@sh\endcsname{\def\PYG@tc##1{\textcolor[rgb]{0.73,0.13,0.13}{##1}}}
\expandafter\def\csname PYG@tok@ow\endcsname{\let\PYG@bf=\textbf\def\PYG@tc##1{\textcolor[rgb]{0.67,0.13,1.00}{##1}}}
\expandafter\def\csname PYG@tok@sx\endcsname{\def\PYG@tc##1{\textcolor[rgb]{0.00,0.50,0.00}{##1}}}
\expandafter\def\csname PYG@tok@bp\endcsname{\def\PYG@tc##1{\textcolor[rgb]{0.00,0.50,0.00}{##1}}}
\expandafter\def\csname PYG@tok@c1\endcsname{\let\PYG@it=\textit\def\PYG@tc##1{\textcolor[rgb]{0.25,0.50,0.50}{##1}}}
\expandafter\def\csname PYG@tok@fm\endcsname{\def\PYG@tc##1{\textcolor[rgb]{0.00,0.00,1.00}{##1}}}
\expandafter\def\csname PYG@tok@o\endcsname{\def\PYG@tc##1{\textcolor[rgb]{0.40,0.40,0.40}{##1}}}
\expandafter\def\csname PYG@tok@kc\endcsname{\let\PYG@bf=\textbf\def\PYG@tc##1{\textcolor[rgb]{0.00,0.50,0.00}{##1}}}
\expandafter\def\csname PYG@tok@c\endcsname{\let\PYG@it=\textit\def\PYG@tc##1{\textcolor[rgb]{0.25,0.50,0.50}{##1}}}
\expandafter\def\csname PYG@tok@mf\endcsname{\def\PYG@tc##1{\textcolor[rgb]{0.40,0.40,0.40}{##1}}}
\expandafter\def\csname PYG@tok@err\endcsname{\def\PYG@bc##1{\setlength{\fboxsep}{0pt}\fcolorbox[rgb]{1.00,0.00,0.00}{1,1,1}{\strut ##1}}}
\expandafter\def\csname PYG@tok@mb\endcsname{\def\PYG@tc##1{\textcolor[rgb]{0.40,0.40,0.40}{##1}}}
\expandafter\def\csname PYG@tok@ss\endcsname{\def\PYG@tc##1{\textcolor[rgb]{0.10,0.09,0.49}{##1}}}
\expandafter\def\csname PYG@tok@sr\endcsname{\def\PYG@tc##1{\textcolor[rgb]{0.73,0.40,0.53}{##1}}}
\expandafter\def\csname PYG@tok@mo\endcsname{\def\PYG@tc##1{\textcolor[rgb]{0.40,0.40,0.40}{##1}}}
\expandafter\def\csname PYG@tok@kd\endcsname{\let\PYG@bf=\textbf\def\PYG@tc##1{\textcolor[rgb]{0.00,0.50,0.00}{##1}}}
\expandafter\def\csname PYG@tok@mi\endcsname{\def\PYG@tc##1{\textcolor[rgb]{0.40,0.40,0.40}{##1}}}
\expandafter\def\csname PYG@tok@kn\endcsname{\let\PYG@bf=\textbf\def\PYG@tc##1{\textcolor[rgb]{0.00,0.50,0.00}{##1}}}
\expandafter\def\csname PYG@tok@cpf\endcsname{\let\PYG@it=\textit\def\PYG@tc##1{\textcolor[rgb]{0.25,0.50,0.50}{##1}}}
\expandafter\def\csname PYG@tok@kr\endcsname{\let\PYG@bf=\textbf\def\PYG@tc##1{\textcolor[rgb]{0.00,0.50,0.00}{##1}}}
\expandafter\def\csname PYG@tok@s\endcsname{\def\PYG@tc##1{\textcolor[rgb]{0.73,0.13,0.13}{##1}}}
\expandafter\def\csname PYG@tok@kp\endcsname{\def\PYG@tc##1{\textcolor[rgb]{0.00,0.50,0.00}{##1}}}
\expandafter\def\csname PYG@tok@w\endcsname{\def\PYG@tc##1{\textcolor[rgb]{0.73,0.73,0.73}{##1}}}
\expandafter\def\csname PYG@tok@kt\endcsname{\def\PYG@tc##1{\textcolor[rgb]{0.69,0.00,0.25}{##1}}}
\expandafter\def\csname PYG@tok@sc\endcsname{\def\PYG@tc##1{\textcolor[rgb]{0.73,0.13,0.13}{##1}}}
\expandafter\def\csname PYG@tok@sb\endcsname{\def\PYG@tc##1{\textcolor[rgb]{0.73,0.13,0.13}{##1}}}
\expandafter\def\csname PYG@tok@sa\endcsname{\def\PYG@tc##1{\textcolor[rgb]{0.73,0.13,0.13}{##1}}}
\expandafter\def\csname PYG@tok@k\endcsname{\let\PYG@bf=\textbf\def\PYG@tc##1{\textcolor[rgb]{0.00,0.50,0.00}{##1}}}
\expandafter\def\csname PYG@tok@se\endcsname{\let\PYG@bf=\textbf\def\PYG@tc##1{\textcolor[rgb]{0.73,0.40,0.13}{##1}}}
\expandafter\def\csname PYG@tok@sd\endcsname{\let\PYG@it=\textit\def\PYG@tc##1{\textcolor[rgb]{0.73,0.13,0.13}{##1}}}

\def\PYGZbs{\char`\\}
\def\PYGZus{\char`\_}
\def\PYGZob{\char`\{}
\def\PYGZcb{\char`\}}

\def\PYGZlt{\char`\<}
\def\PYGZgt{\char`\>}
\def\PYGZsh{\char`\#}

\def\PYGZdl{\char`\$}
\def\PYGZhy{\char`\-}
\def\PYGZsq{\char`\'}
\def\PYGZdq{\char`\"}

\makeatother

\copyrightyear{2022}
\acmYear{2022}
\setcopyright{acmcopyright}
\acmConference[CODASPY '22]{Proceedings of the Twelveth ACM Conference on Data and Application Security and Privacy}{April 24--27, 2022}{Baltimore, MD, USA}
\acmBooktitle{Proceedings of the Twelveth ACM Conference on Data and Application Security and Privacy (CODASPY '22), April 24--27, 2022, Baltimore, MD, USA}
\acmPrice{15.00}
\acmDOI{10.1145/3508398.3511525}
\acmISBN{978-1-4503-9220-4/22/04}

\SetWatermarkText{Submitted}

\begin{document}
\StopCensoring

\title[Hardening with Scapolite]{Hardening with Scapolite: a DevOps-based Approach for Improved Authoring and Testing of Security-Configuration Guides in Large-Scale Organizations}

\author{Patrick Stöckle}
\email{patrick.stoeckle@tum.de}
\orcid{0000-0003-0193-5871}

\affiliation{%
    \institution{Technical University of Munich (TUM)}
    \city{Munich}
    \country{Germany}
}

\author{Ionuț Pruteanu}
\email{ionut.pruteanu@siemens.com}

\affiliation{%
    \institution{Siemens AG}
    \city{Bucharest}
    \country{Romania}
}

\author{Bernd Grobauer}
\email{bernd.grobauer@siemens.com}

\affiliation{%
  \institution{Siemens AG}
  \city{Munich}
  \country{Germany}
}

\author{Alexander Pretschner}
\email{alexander.pretschner@tum.de}
\orcid{0000-0002-5573-1201}
\affiliation{%
    \institution{Technical University of Munich (TUM)}
    \city{Munich}
    \country{Germany}
}

\begin{abstract}
\textsc{Fullpaper}\footnote{We submitted this article as a full-length paper.
Unfortunately, the CODASPY Program Committee decided that our paper can only be accepted in the tool track.
Thus, the published version only consists of 6 pages.}
Security Hardening is the process of configuring IT systems to ensure the security of the systems' components and data they process or store.
In many cases, so-called security-configuration guides are used as a basis for security hardening.
These guides describe secure configuration settings for components such as operating systems and standard applications.
Rigorous testing of security-configuration guides and automated mechanisms for their implementation and validation are necessary since erroneous implementations or checks of hardening guides may severely impact systems' security and functionality.
At \censor{Siemens}, centrally maintained security-configuration guides carry machine-readable information specifying both the implementation and validation of each required configuration step.
The guides are maintained within \emph{git} repositories;
automated pipelines generate the artifacts for implementation and checking, e.g., PowerShell scripts for Windows, and carry out testing of these artifacts on AWS images.
This paper describes our experiences with our DevOps-inspired approach for authoring, maintaining, and testing security-configuration guides.
We want to share these experiences to help other organizations with their security hardening and, thus, increase their systems' security.
\end{abstract}


\begin{teaserfigure}
\includegraphics[width=\columnwidth]{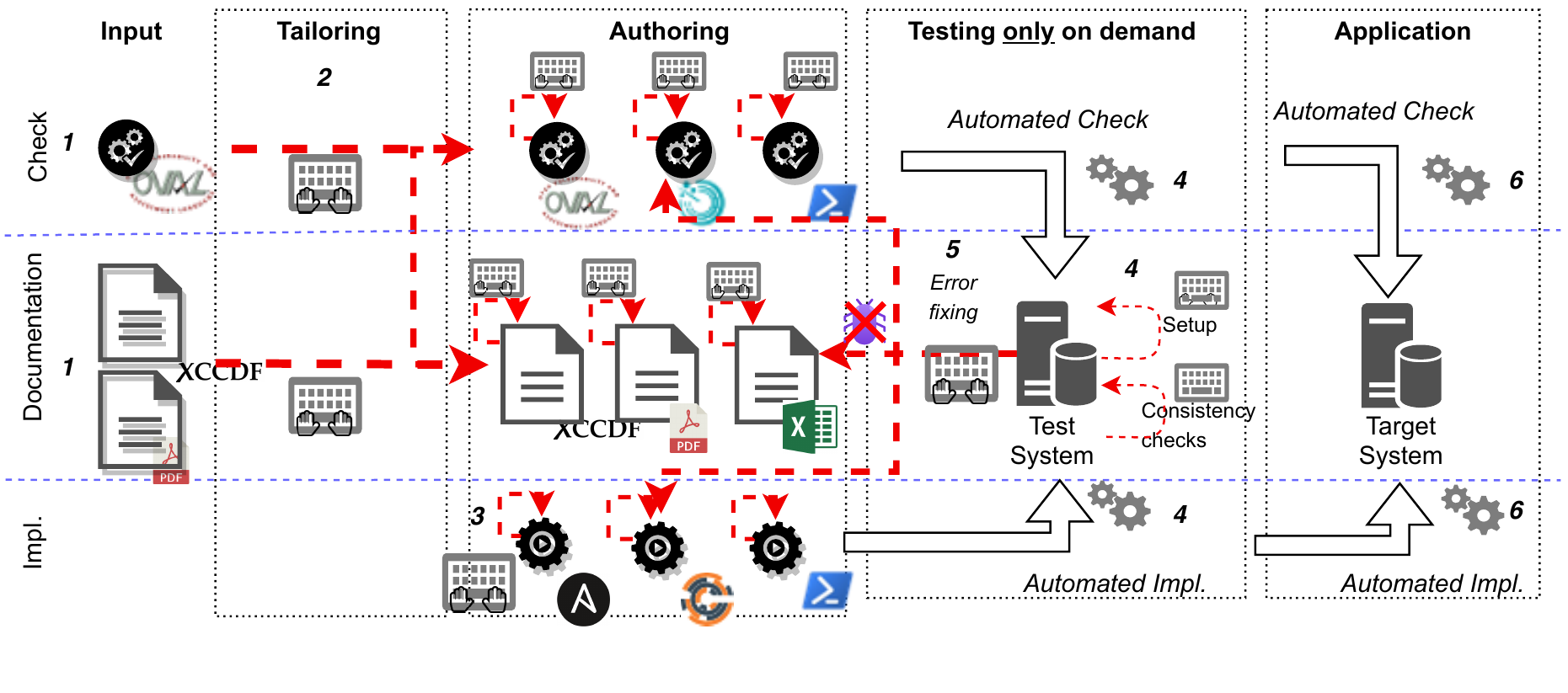}
\caption{Typical process of security hardening.
Dotted arrows represent manual tasks.
Every arrow within the box is a task the administrators execute to harden the system.}
\label{fig:hardeningProcess}
\end{teaserfigure}

\maketitle

\section{Introduction}
\label{sec:intro}

\begin{figure*}[t]
\includegraphics[width=2\columnwidth]{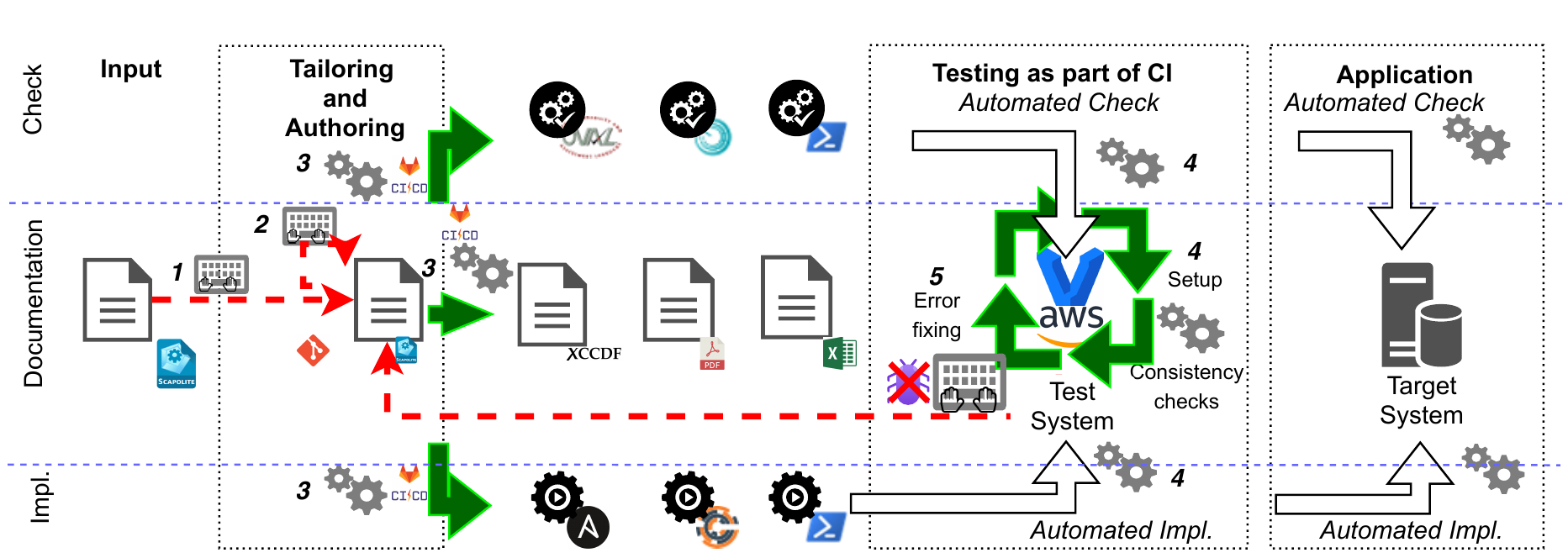}
\caption{Improved process of security hardening. The green arrows represent activities that have been automated.}
\label{fig:newHardeningProcess}
\end{figure*}

Insecure configurations of operating systems and applications are known to be both common and detrimental to cybersecurity~\cite{Dietrich:2018:ISO:3243734.3243794,Tang:2015:HCM:2815400.2815401,Continella:2018:THB:3274694.3274736,Jha2017,Xu:2015:SAT:2775083.2791577}.
Organizations, therefore, need to identify the security-relevant configuration settings of the used software, determine the secure value or a set of secure values for each setting, and ensure that they configure each instance of the software used within their organization accordingly.
This process is called security-configuration hardening and is part of the general security hardening of an organization's infrastructure.
Security hardening is a continuous process rather than a one-time-only task since the IT infrastructure, the threat environment, insights about (in)secure configurations, et cetera are constantly in flux.

Organizations such as the Center for Internet Security (CIS) or the Defense Information Systems Agency (DISA) provide publicly available security-configuration guides (also called benchmarks, guidelines, or baselines) for various software components, e.g., operating systems like Windows 10, web servers like NGINX, or email clients like Outlook.
These guides consist of rules, and each rule states which values should be used for a configuration setting relevant for security;
some of these guides consist of more than 350 rules.
Benchmarks written in the SCAP~\cite{scap_v1_3} standard often contain machine-readable definitions of \emph{checks}, whereas mechanisms for \emph{implementing} the required settings are usually either provided separately or not at all.
The usual security-configuration hardening process, which is based on such public guides, contains many manual steps that are both inefficient and error-prone.
Most of the time, we need to adapt the external guides for our target infrastructure by modifying specific settings, removing some rules, and adding others.
This problem is intensified by the fact that these adaptions have to be replicated and kept consistent for each implementation, such as scripts (e.g., Bash or PowerShell), Infrastructure as Code (IaC) approaches (e.g.,  Ansible or Chef), et cetera, and for each check mechanism.

\subsection{Problems of the Current Security Hardening Process}

Figure \ref{fig:hardeningProcess} illustrates the usual security hardening process;
the numbers in the figure refer to the following steps:

\begin{enumerate}
\item Input is an external guide, usually in the SCAP standard:
The human-readable parts are defined in the \emph{eXtensible Configuration Checklist Format} (XCCDF) with machine-readable checks in the \emph{Open Vulnerability and Assessment Language} (OVAL).
\item XCCDF offers a mechanism for tailoring the guide, e.g., configure changes via so-called profiles.
The profiles are also reflected in the OVAL-based checks.
\item Because machine-readable implementation mechanisms are not part of these guides (exception: ComplianceAsCode, discussed below), we must either manually develop implementation mechanisms or adjust them if we can re-use existing mechanisms.
Since larger organizations may use several different implementation mechanisms, we may need to re-apply the same changes numerous times.
\item Before applying the implementation mechanisms to and using the check mechanisms for production systems, we must test both of them:
Erroneous implementation and checking of security configurations may severely impact the security and functionality of systems.
Because security-configuration guides are used for many target systems (different operating systems and applications, different releases, different tailorings, et cetera), we must manage a corresponding multitude of test systems.
\item Feedback about problems, e.g., faulty implementations or checks, might introduce changes for one or several implementation/check mechanisms.
\item Finally, the tailored and tested security guides can be applied to production systems.
If problems are detected in productive use or a new version of a guide is published, the whole process restarts.
\end{enumerate}

The repetition of these \textbf{manual} steps increases the risk of introducing \textbf{errors} and, thus, the risk of \textbf{insecure systems}.
Therefore, we identified the following \emph{challenges} for improving the security hardening:
\begin{itemize}
    \item Remove superfluous complexity in the security hardening process resulting from unnecessary manual steps and scattered information.
    \item Establish automatic quality assurance for the security-configuration guides to find errors earlier and easier.
\end{itemize}
\subsection{Our Approach: Improved Authoring, Artifact Generation, and Automated Testing}
Our solution to these challenges is twofold.
First, we present our improved configuration hardening approach that focuses on automation to remove error-prone manual steps.
Second, we present our approach on automatic testing of security-configuration guides to detect errors as soon as possible.

Figure~\ref{fig:newHardeningProcess} shows our improved security hardening process;
again, the numbers refer to the steps below:
\begin{enumerate}
\item We manage security-configuration guides in a dedicated YAML-based format called \emph{Scapolite}, which we keep under version control.
Further, we enrich the format with machine-readable information about configuration requirements.
Ideally, both implementation and check mechanisms can be automatically derived.
Thus, we keep information about the check, implementation, metadata, and documentation, e.g., human-readable descriptions about the requirements, the rationale, et cetera, at a single location.\footnote{External guides in SCAP can be automatically converted into Scapolite.
Adding machine-readable information from which implementations and checks can be derived requires, of course, manual effort, but such  effort would be necessary for generating separate implementation mechanism, as well. Furthermore, for some use-cases, semi-automated mechanisms for deriving machine-readable information from human-readable text exist \cite{10.1145/3324884.3416540}.}
  
\item Tailoring to different use-cases in Scapolite works similarly as in SCAP:
We can define profiles for the individual use cases and create per-use-case modifications.

\item From this single source, i.e., the machine-readable information from 1), we automatically generate the required artifacts for implementing/checking the guides.
\item Creation of the required test systems as virtual machines, applying the implementations/checks to these systems, and collecting the test results is carried out automatically as a part of a DevOps pipeline. 

\item Because the implementations/checks are generated automatically, we can fix detected problems with a single change either in the Scapolite document defining the guide or a bug-fix in the transformation system, rather than changing in several different artifacts. 
\end{enumerate}

\begin{figure}[t]
\includegraphics[width=\columnwidth]{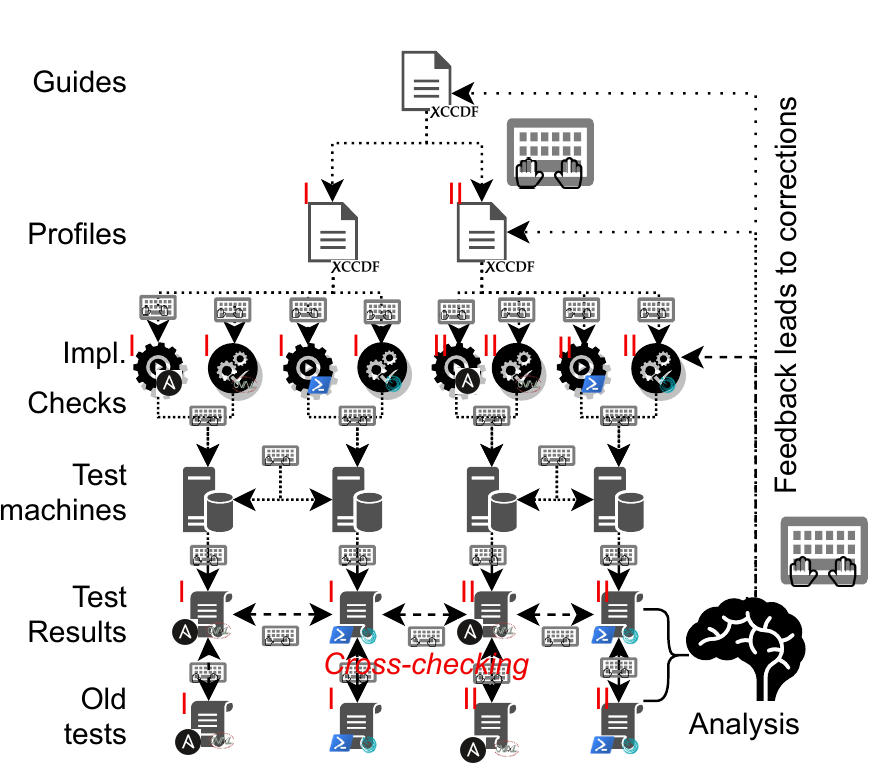}
\caption{Regular execution of tests in a security hardening process.
Dotted lines denote manual tasks. Every arrow within the box is a task the administrators execute to harden the system.}
\label{fig:oldtestingprocess}
\end{figure}

\subsection{Contributions}

Our contributions to the field of security hardening are:
\begin{itemize}[leftmargin=0.5em]
    \item By pulling information required for generating both implementation and check mechanisms as machine-readable information into our security-configuration guides, we manage to restrict manual changes/corrections to a single location, thus reducing errors and increasing efficiency.
    \item We show how to operate a DevOps/Continuous Integration-inspired approach of authoring and maintaining security-configuration guides.
    In our approach, changes in the guides trigger automated tests without human involvement in the execution of the tests, collection of test results, and correlation of test data with expected results. 
\end{itemize}

\begin{figure}[t]
\includegraphics[width=\columnwidth]{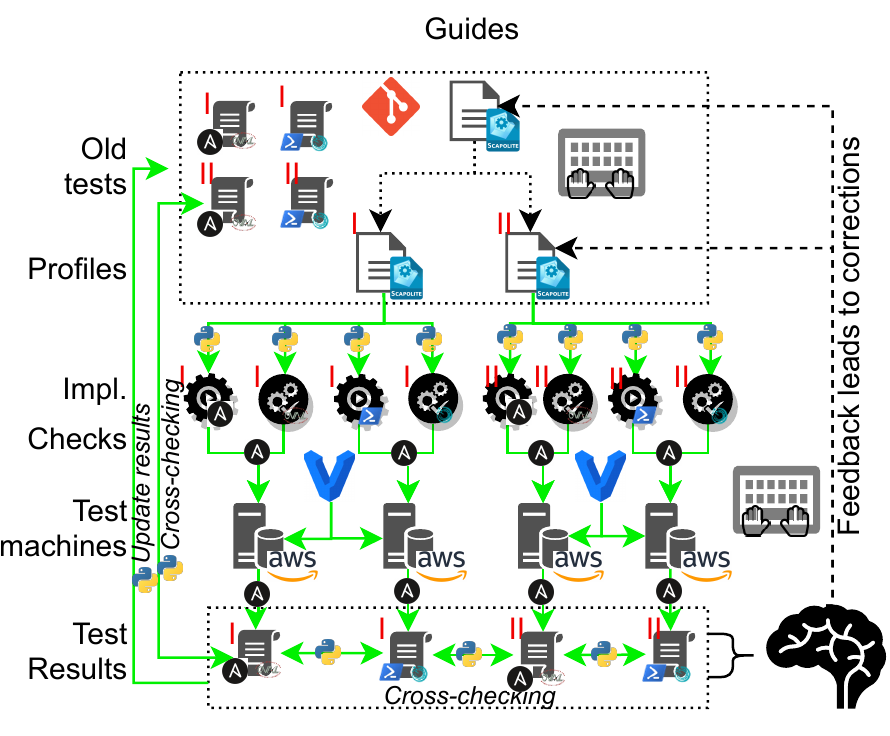}
\caption{State-of-the-art execution of tests in a security hardening process.
The green arrows denote steps that are now automated.}
\label{fig:newtestingprocess}
\end{figure}

The latter point deserves a closer examination:
As explained above, security-configuration mechanisms are affected by the combinatorial explosion of test cases, requiring many test systems and test runs.
Figure~\ref{fig:oldtestingprocess} illustrates the approach without the DevOps:
a single test already requires a substantial manual effort that must be multiplied by the number of test systems/test cases;
when we detect problems, we have to fix them at several locations.
In contrast, Figure~\ref{fig:newtestingprocess} illustrates the level of automation of our approach.

Our experiences of handling multiple security-configuration guides with multiple profiles authored/maintained using version control and DevOps pipelines within an industrial context show that an approach that combines machine-readable information required for implementing and checking security-configuration requirements is not only feasible but provides enormous benefits.
Errors are reduced, and the efficiency and the effectiveness of an organization’s security-configuration hardening program are raised.
Thus, we tackle two of the major causes for insecure configurations:
erroneous application and ineffective or incomplete application of secure configurations.

\section{Our approach to security hardening}
\label{sec:approach-to-security-hardening}

Everyone who has published security-configuration guidelines to their organization, e.g., a document specifying the required security settings for a \emph{Windows} or \emph{Linux} server system, will
be familiar with the demand for means that allow automated implementation and validation of these settings.
Especially in the case of operating systems, for which the number
of relevant settings is over 350, publishing a guide without providing automated mechanisms
is both inefficient and ineffective:
\begin{itemize}[leftmargin=0.5em]
    \item multiple persons/groups in the constituency work in parallel on creating implementation/validation mechanisms;
    \item the manual transcription of required settings into an implementation mechanism or a fully manual implementation will lead to errors and omissions;
    \item some constituency members will deem the task of implementation as too arduous, costly, or time-consuming and not bother with it at all.
\end{itemize}
The SCAP~\cite{scap_v1_3} format family defines the state of the art for providing automated mechanisms along with a security-configuration guide.
We can use the SCAP formats to augment human-readable information with machine-readable checks, usually specified in OVAL~\cite{oval_documentation}.
In almost all cases, however, automated implementation mechanisms are maintained separately:
both CIS and DISA provide Windows backup files containing the required settings, which need to be maintained manually -- a cumbersome and error-prone process, as outlined above.
The notable exception is the \textit{ComplianceAsCode}~\footnote{\url{https://github.com/ComplianceAsCode/content}} project that provides little scripts or Ansible playbooks for many settings.
ComplianceAsCode includes the scripts in the resulting SCAP content such that tools can use them to carry out the implementation steps.
At \censor{Siemens}, we take a similar approach to ComplianceAsCode.
However, we try to operate at a higher level of abstraction -- where possible -- by specifying the desired configurations in a machine-readable form such that we can derive both implementation and verification mechanisms from it.
We combine this with a rigorous ``DevOps''-approach for authoring and maintaining security guides:
we use DevOps pipelines for both automated derivation and test of implementation and validation mechanisms.
In the following, we will briefly outline our approach towards the abstract specification of security-configuration requirements and their automated transformation into implementations and checks.

\subsection{The Scapolite Format}

The starting point of our work was the definition of a format called
``Scapolite,'' which encompasses the relevant features of SCAP but additionally provides

\begin{enumerate}
    \item a form that can be created/maintained as text-files under version control (cf. above comment on changes in rules).
    \item generalizations and additional extension points to support a broader range of use cases.
    \item fields for tracking of document maintenance data such as change history information per configuration requirement.
\end{enumerate}

Similar to other projects~\cite{ansible_documentation,opencontrol} that require a ``human read- and writable'' format for creating and maintaining structured information, we chose YAML\footnote{\url{https://yaml.org/}} as a basis for Scapolite.
Further, we combined YAML with Markdown\footnote{
\url{https://daringfireball.net/projects/markdown/}
} as a markup language for structuring human-readable content.
We do not argue that SCAP and its XML formats OVAL, XCCDF, et cetera are not human-readable, but our experience from working with guide authors at \censor{Siemens} shows that they are more motivated to write  guides in a YAML/Markdown than in an XML format.

Listing~\ref{lst:basic_scapolite} shows a minimal example of Scapolite; the highlighted lines contain the human-readable description of how to implement the required setting.

\begin{listing}[!htb]
\begin{Verbatim}[commandchars=\\\{\}]
\PYGZhy{}\PYGZhy{}\PYGZhy{}
scapolite:
    class: rule
    version: \PYGZsq{}0.51\PYGZsq{}
id: BL942\PYGZhy{}1101
id\PYGZus{}namespace: org.scapolite.example
title: Configure use of passwords for removable data drives
rule: \PYGZlt{}see below\PYGZgt{}
implementations:
    \PYG{k}{\PYGZhy{}} relative\PYGZus{}id: \PYGZsq{}01\PYGZsq{}
    description: \PYGZlt{}see below\PYGZgt{}
history:
    \PYG{k}{\PYGZhy{}} version: \PYGZsq{}1.0\PYGZsq{}
    action: created
    description: Added so as to mitigate risk SR\PYGZhy{}2018\PYGZhy{}0144.
\PYGZhy{}\PYGZhy{}\PYGZhy{}
\PYG{g+gu}{\PYGZsh{}\PYGZsh{}} /rule
Enable the setting \PYGZsq{}Configure use of passwords for removable
data drives\PYGZsq{} and set the options as follows:
    \PYG{k}{*}  Select \PYG{l+s+sb}{`Require password complexity`}
    \PYG{k}{*}  Set the option \PYGZsq{}Minimum password length for removable data drive` to \PYG{l+s+sb}{`15`}.
\PYG{g+gu}{\PYGZsh{}\PYGZsh{}} /implementations/0/description
To set the protection level to the desired state, enable the policy
\PYG{l+s+sb}{`Computer Configuration\PYGZbs{}...\PYGZbs{}Configure use of passwords for removable data drives`}
and set the options as specified above in the rule.
\end{Verbatim}

\caption{A very basic example of a rule in Scapolite.
Lines referenced in the text are marked in blue.
We shortened the policy path to keep the file concise.}
\label{lst:basic_scapolite}
\end{listing}

\subsection{Adding Machine-Readable Automations}
\label{sec:machine-readable-automations}

The setting prescribed by the rule in Listing~\ref{lst:basic_scapolite} concerns a Windows policy setting, specified via (1) a policy path and (2) the required \textit{policy value}.
We, therefore, augment the Scapolite rule object shown in listing~\ref{lst:basic_scapolite} with a so-called \emph{automation} structure:
the \texttt{implementation} section of that object has an optional keyword \texttt{automations} under which we can add a list of such \emph{automation} structures.
Listing~\ref{lst:gpo_automation} shows the required automation  structure for this particular rule.
Line 2 contains the policy path;
starting with line 4, one can see the required values.
In addition to the policy path and the values, in lines 7-10, we also specify constraints for compliance checking:
obviously, a password length $ > 15 $ would also be compliant.

\begin{listing}[!htb]
\begin{Verbatim}[commandchars=\\\{\}]
\PYG{n+nt}{system}\PYG{p}{:} \PYG{l+lScalar+lScalarPlain}{org.scapolite.implementation.win\PYGZus{}gpo}
\PYG{n+nt}{ui\PYGZus{}path}\PYG{p}{:} \PYG{l+lScalar+lScalarPlain}{Computer Configuration\PYGZbs{}...\PYGZbs{}Configure use of passwords for removable data drives}
\PYG{n+nt}{value}\PYG{p}{:}
    \PYG{n+nt}{main\PYGZus{}setting}\PYG{p}{:} \PYG{l+lScalar+lScalarPlain}{Enabled}
    \PYG{n+nt}{Configure password complexity for removable data drives}\PYG{p}{:} \PYG{l+lScalar+lScalarPlain}{Require password complexity}
    \PYG{n+nt}{Minimum password length for removable data drive}\PYG{p}{:} \PYG{l+lScalar+lScalarPlain}{15}
    \PYG{n+nt}{constraints}\PYG{p}{:}
    \PYG{n+nt}{Minimum password length for removable data drive}\PYG{p}{:}
        \PYG{n+nt}{min}\PYG{p}{:} \PYG{l+lScalar+lScalarPlain}{15}
\end{Verbatim}

\caption{Windows-policy automation specifying a policy path, value(s) and constraints for compliance checking}
\label{lst:gpo_automation}
\end{listing}

We can configure Windows policies via a GUI interface, which allows the user to choose the desired values for each existing policy path.
For a programmatic implementation, however, an intermediate step is necessary.
In the case of this particular policy, we must set a specific key-value pair in the registry.
We have, therefore, implemented an automated transformation of the policy-based specification to a registry-based automation (similar transformations exist for other  ``low-level'' mechanisms required for other Windows policies).

\subsection{Transforming Automations}
\label{sec:transforming-automations}

Listing~\ref{lst:registry_automation_example} provides the result of carrying out this transformation for the automation in Listing~\ref{lst:gpo_automation}:
we must set three registry keys;
the first key signifies that the setting is enabled;
the second specifies that the requirements on password complexity are active;
the third contains the minimum password length.

\begin{listing}[!htb]
\begin{Verbatim}[commandchars=\\\{\}]
\PYG{n+nt}{system}\PYG{p}{:} \PYG{l+lScalar+lScalarPlain}{org.scapolite.automation.compound}
\PYG{n+nt}{automations}\PYG{p}{:}
    \PYG{p+pIndicator}{\PYGZhy{}} \PYG{n+nt}{system}\PYG{p}{:} \PYG{l+lScalar+lScalarPlain}{org.scapolite.implementation.windows\PYGZus{}registry}
    \PYG{n+nt}{config}\PYG{p}{:} \PYG{l+lScalar+lScalarPlain}{Computer}
    \PYG{n+nt}{registry\PYGZus{}key}\PYG{p}{:} \PYG{l+lScalar+lScalarPlain}{Software\PYGZbs{}Policies\PYGZbs{}Microsoft\PYGZbs{}FVE}
    \PYG{n+nt}{value\PYGZus{}name}\PYG{p}{:} \PYG{l+lScalar+lScalarPlain}{RDVPassphrase}
    \PYG{n+nt}{action}\PYG{p}{:} \PYG{l+lScalar+lScalarPlain}{DWORD:1}
    \PYG{p+pIndicator}{\PYGZhy{}} \PYG{n+nt}{system}\PYG{p}{:} \PYG{l+lScalar+lScalarPlain}{org.scapolite.implementation.windows\PYGZus{}registry}
    \PYG{n+nt}{config}\PYG{p}{:} \PYG{l+lScalar+lScalarPlain}{Computer}
    \PYG{n+nt}{registry\PYGZus{}key}\PYG{p}{:} \PYG{l+lScalar+lScalarPlain}{Software\PYGZbs{}Policies\PYGZbs{}Microsoft\PYGZbs{}FVE}
    \PYG{n+nt}{value\PYGZus{}name}\PYG{p}{:} \PYG{l+lScalar+lScalarPlain}{RDVPassphraseComplexity}
    \PYG{n+nt}{action}\PYG{p}{:} \PYG{l+lScalar+lScalarPlain}{DWORD:1}
    \PYG{p+pIndicator}{\PYGZhy{}} \PYG{n+nt}{system}\PYG{p}{:} \PYG{l+lScalar+lScalarPlain}{org.scapolite.implementation.windows\PYGZus{}registry}
    \PYG{n+nt}{config}\PYG{p}{:} \PYG{l+lScalar+lScalarPlain}{Computer}
    \PYG{n+nt}{registry\PYGZus{}key}\PYG{p}{:} \PYG{l+lScalar+lScalarPlain}{Software\PYGZbs{}Policies\PYGZbs{}Microsoft\PYGZbs{}FVE}
    \PYG{n+nt}{value\PYGZus{}name}\PYG{p}{:} \PYG{l+lScalar+lScalarPlain}{RDVPassphraseLength}
    \PYG{n+nt}{action}\PYG{p}{:} \PYG{l+lScalar+lScalarPlain}{DWORD:15}
    \PYG{n+nt}{constraints}\PYG{p}{:}
        \PYG{n+nt}{min}\PYG{p}{:} \PYG{l+lScalar+lScalarPlain}{15}
\end{Verbatim}

\caption{Example of the Windows Registry automations generated from Listing~\ref{lst:gpo_automation}}
\label{lst:registry_automation_example}
\end{listing}



Ideally, all security requirements should be specified as abstractly as possible and then be transformed automatically into mechanisms for implementation and checking.
However, if we cannot find a suitable abstraction level, we must include code in a suitable scripting language.
For expressing checks, we can at least regain some abstraction via a generic method for expressing the expected output of check-scripts to keep the scripts included as ``script automation'' in the Scapolite document as concise as possible.
Listing~\ref{lst:script_example} shows an example of a check for the requirement that all mounted volumes larger than 1GB should use the NTFS file system.
Lines~6-8 specify the expected output:
the script in line 3 returns a list of information objects, each of which must carry the key-value pair \verb+FileSystemType:NTFS+.

\begin{listing}[!htb]
\begin{Verbatim}[commandchars=\\\{\}]
\PYG{n+nt}{system}\PYG{p}{:} \PYG{l+lScalar+lScalarPlain}{org.scapolite.automation.script}
\PYG{n+nt}{script}\PYG{p}{:} \PYG{p+pIndicator}{|}
    \PYG{n+no}{Get\PYGZhy{}Volume | Select Size, FileSystemType | Where \PYGZob{}\PYGZdl{}\PYGZus{}.Size \PYGZhy{}gt 1GB\PYGZcb{}}
\PYG{n+nt}{expected}\PYG{p}{:}
    \PYG{n+nt}{output\PYGZus{}processor}\PYG{p}{:} \PYG{l+lScalar+lScalarPlain}{Format\PYGZhy{}List}
    \PYG{n+nt}{each\PYGZus{}item}\PYG{p}{:}
        \PYG{n+nt}{key}\PYG{p}{:} \PYG{l+lScalar+lScalarPlain}{FileSystemType}
        \PYG{n+nt}{equal\PYGZus{}to}\PYG{p}{:} \PYG{l+lScalar+lScalarPlain}{NTFS}
\end{Verbatim}
\caption{Example of a script-based automation for checking that all drives larger than 1GB use NTFS as their file system type.}
\label{lst:script_example}
\end{listing}

\subsection{Producing Code and Other Artifacts}
\label{section:generate_code}


With (1) the machine-readable specifications of what needs to be implemented/checked and (2) the associated transformation mechanisms, we can generate artifacts that the system administrators can use to carry out the rule's implementation and check.
The higher our level of abstraction, the more options we have regarding the target implementation or check mechanism for which we generate these artifacts.
Obviously, if the automations contain code for a specific script engine, we must generate artifacts for each of these engines or an execution system that can execute this type of script.

For this article, we continue the example regarding Windows.
For security-configuration guides targeting Windows, we generate a set of PowerShell commandlets together with a JSON file containing for each rule the necessary data used by the PowerShell commandlets to implement or check the rule.
Before the scripts implement a rule, they store as backup each setting's current value;
Thus, we can roll back every implemented rule.

As an example for a different \textit{target} of our transformations, Listing~\ref{lst:registry_automation_example_oval} shows the result of a transformation from Listing~\ref{lst:registry_automation_example} into an OVAL check.
This particular transformation might look straightforward, but even simple checks can get complicated when expressed in OVAL;
combined with the verbose XML structure of OVAL and its many cross-references, generating OVAL was a prime use case for our code generation.

Our improved approach to security hardening has several advantages:
First, it concentrates all information of a single rule in one place and reduces the risk of inconsistencies.
Second, the transformations replace many manual steps and thus significantly reduce the risk of errors.

\begin{listing}[!htb]
\begin{Verbatim}[commandchars=\\\{\}]
\PYG{n+nt}{\PYGZlt{}criteria} \PYG{n+na}{negate=}\PYG{l+s}{\PYGZdq{}false\PYGZdq{}} \PYG{n+na}{operator=}\PYG{l+s}{\PYGZdq{}AND\PYGZdq{}}\PYG{n+nt}{\PYGZgt{}}
    \PYG{n+nt}{\PYGZlt{}criteria} \PYG{n+na}{negate=}\PYG{l+s}{\PYGZdq{}false\PYGZdq{}} \PYG{n+na}{operator=}\PYG{l+s}{\PYGZdq{}AND\PYGZdq{}}\PYG{n+nt}{\PYGZgt{}}
    \PYG{n+nt}{\PYGZlt{}criterion} \PYG{n+na}{negate=}\PYG{l+s}{\PYGZdq{}false\PYGZdq{}} \PYG{n+na}{test\PYGZus{}ref=}\PYG{l+s}{\PYGZdq{}oval:tst:105650\PYGZdq{}}\PYG{n+nt}{\PYGZgt{}}
        \PYG{n+nt}{\PYGZlt{}win:registry\PYGZus{}test} \PYG{n+na}{check=}\PYG{l+s}{\PYGZdq{}all\PYGZdq{}} \PYG{n+na}{check\PYGZus{}existence=}\PYG{l+s}{\PYGZdq{}at\PYGZus{}least\PYGZus{}one\PYGZus{}exists\PYGZdq{}}  \PYG{n+na}{id=}\PYG{l+s}{\PYGZdq{}oval:tst:105650\PYGZdq{}} \PYG{n+na}{version=}\PYG{l+s}{\PYGZdq{}1\PYGZdq{}}\PYG{n+nt}{\PYGZgt{}}
        \PYG{n+nt}{\PYGZlt{}win:registry\PYGZus{}object} \PYG{n+na}{id=}\PYG{l+s}{\PYGZdq{}oval:obj:105650\PYGZdq{}} \PYG{n+na}{version=}\PYG{l+s}{\PYGZdq{}1\PYGZdq{}}\PYG{n+nt}{\PYGZgt{}}
            \PYG{n+nt}{\PYGZlt{}win:hive} \PYG{n+na}{datatype=}\PYG{l+s}{\PYGZdq{}string\PYGZdq{}} \PYG{n+na}{operation=}\PYG{l+s}{\PYGZdq{}equals\PYGZdq{}}\PYG{n+nt}{\PYGZgt{}}
            HKEY\PYGZus{}LOCAL\PYGZus{}MACHINE
            \PYG{n+nt}{\PYGZlt{}/win:hive\PYGZgt{}}
            \PYG{n+nt}{\PYGZlt{}win:key} \PYG{n+na}{datatype=}\PYG{l+s}{\PYGZdq{}string\PYGZdq{}} \PYG{n+na}{operation=}\PYG{l+s}{\PYGZdq{}case insensitive equals\PYGZdq{}}\PYG{n+nt}{\PYGZgt{}}
            Software\PYGZbs{}Policies\PYGZbs{}Microsoft\PYGZbs{}FVE
            \PYG{n+nt}{\PYGZlt{}/win:key\PYGZgt{}}
            \PYG{n+nt}{\PYGZlt{}win:name} \PYG{n+na}{datatype=}\PYG{l+s}{\PYGZdq{}string\PYGZdq{}} \PYG{n+na}{operation=}\PYG{l+s}{\PYGZdq{}equals\PYGZdq{}}\PYG{n+nt}{\PYGZgt{}}
            RDVPassphrase
            \PYG{n+nt}{\PYGZlt{}/win:name\PYGZgt{}}
        \PYG{n+nt}{\PYGZlt{}/win:registry\PYGZus{}object\PYGZgt{}}
        \PYG{n+nt}{\PYGZlt{}win:registry\PYGZus{}state} \PYG{n+na}{id=}\PYG{l+s}{\PYGZdq{}oval:ste:105650\PYGZdq{}} \PYG{n+na}{version=}\PYG{l+s}{\PYGZdq{}1\PYGZdq{}}\PYG{n+nt}{\PYGZgt{}}
            \PYG{n+nt}{\PYGZlt{}win:type} \PYG{n+na}{datatype=}\PYG{l+s}{\PYGZdq{}string\PYGZdq{}} \PYG{n+na}{operation=}\PYG{l+s}{\PYGZdq{}equals\PYGZdq{}}\PYG{n+nt}{\PYGZgt{}}
            reg\PYGZus{}dword
            \PYG{n+nt}{\PYGZlt{}/win:type\PYGZgt{}}
            \PYG{n+nt}{\PYGZlt{}win:value} \PYG{n+na}{datatype=}\PYG{l+s}{\PYGZdq{}int\PYGZdq{}} \PYG{n+na}{entity\PYGZus{}check=}\PYG{l+s}{\PYGZdq{}all\PYGZdq{}} \PYG{n+na}{operation=}\PYG{l+s}{\PYGZdq{}equals\PYGZdq{}}\PYG{n+nt}{\PYGZgt{}}
            1
            \PYG{n+nt}{\PYGZlt{}/win:value\PYGZgt{}}
        \PYG{n+nt}{\PYGZlt{}/win:registry\PYGZus{}state\PYGZgt{}}
        \PYG{n+nt}{\PYGZlt{}/win:registry\PYGZus{}test\PYGZgt{}}
    \PYG{n+nt}{\PYGZlt{}/criterion\PYGZgt{}}
    \PYG{n+nt}{\PYGZlt{}/criteria\PYGZgt{}}
    ...
\PYG{n+nt}{\PYGZlt{}/criteria\PYGZgt{}}
\end{Verbatim}

\caption{Parts of an OVAL check (nested for better readability) generated from listing~\ref{lst:registry_automation_example}.
Shown is the part of the check that considers the first of the three registry keys.}
\label{lst:registry_automation_example_oval}
\end{listing}

\section{The need for automated testing}
\label{sec:need_for_testing}

Having explained how we specify security-configuration requirements and transform these specifications into artefacts for implementation and checking, we move to motivating the need for extensive test automation as part of our maintenance and release process in the following section.

\subsection{Maintenance and Release Process}
\label{sec:process}

Our workflow in authoring, maintaining, and releasing security-configuration baselines is as follows:

\begin{enumerate}
  \item Authors write security-configuration guides using Scapolite.
  The Scapolite files are kept under version control at \texttt{code.\censor{siemens}.com}, an internal \emph{GitLab} instance.
  \item We use GitLab pipelines to automatically transform the machine-readable automations into artifacts for implementation and check, i.e.,  in the Windows case, we generate JSON files and PowerShell scripts.
  During the development or maintenance of a guide, the authors use these guides for testing purposes.
  \item Once we release a guide, \censor{Siemens}'s security-regulation portal called \emph{\censor{SFeRA}} generates human-readable versions (web view, PDF, XLSX, etc.) directly from the Scapolite sources located at \texttt{code.\censor{siemens}.com}. 
  \item The pipeline-based transformation mechanism is triggered for the released version of the Scapolite sources, and we provide the resulting artifacts to users via dedicated GitLab repositories.
\end{enumerate}

In a parallel process, we maintain the technological basis of this process and develop it further, namely:

\begin{enumerate}
  \item libraries for creating and manipulating Scapolite content, e.g., imports from SCAP, methods for enriching existing Scapolite content with additional information, et cetera;
  \item libraries for transforming abstract machine-readable automations into more concrete automations, e.g., transforming a Windows policy requirement into registry key settings (cf. Section~\ref{sec:transforming-automations});
  \item libraries for further transformation into code or other artifacts (cf. Section~\ref{section:generate_code})
\end{enumerate}

\subsection{Combinatorial Explosion of Needed Test Cases}
\label{sec:multitude_of_test_cases}

Before describing the test requirements during the creation/maintenance of security-configuration guides, it is worthwhile to consider the number of test cases for a given guide.

Usually, we write security-configuration guides to serve different use-cases with the same guide.
We normally specify different security levels, where specific rules only apply to particular levels or rules are modified according to the security level.
For example, a lower password length may be required for standard systems, whereas we specify a longer password length for high-security systems or add a rule mandating two-factor authentication.

Also, frequently, we differentiate between other use-case variants such as client and server systems.
Thus, a scheme for defining system criticality or sensitivity in three levels for an OS, i.e., low, medium, high, as well as for two roles, i.e., client and server, will lead to 6 test cases.
For \censor{Siemens's} Enterprise IT, we use a criticality schemes which (in theory) can lead to 27 different possible criticality levels.

Finally, a single security-configuration guide may apply to several releases of its target, e.g., Windows releases (1809, et cetera), or different editions or flavors of the target, e.g., CentOS vs.
RHEL.

Thus, we see that testing of security guides suffers from a substantial combinatorial explosion problem.
We know mitigation strategies, e.g., containerization, to provide a controlled execution environment to remove the variability;
we cannot apply them since security-configuration guides strive to be applicable as widely as possible.

\subsection{Test Requirements during Guide Creation}
\label{sec:test_requirements_during_creation}

Creating a security-configuration guide is an iterative process between writing the guide and testing the guide's implementation.
The author, therefore, requires a test environment, usually in the form of one or more virtual images on which the target of the baseline is installed.

Manual creation/maintenance of such a test environment, as well as the manual execution of the tests, is a tremendous overhead:
we must start/reset the virtual image, generate the artifacts, transfer them to the image, and execute the artifacts; usually, we execute this process several times for implementing and checking rules for different use-cases. In the end, we must collect the test results and prepare them for the manual analysis.

The efficient creation of security-configuration guides, therefore, is impossible without automated testing.

\subsection{Test Requirements During Guide Maintenance}
\label{sec:test_requirements_during_maintenance}

Automated testing also is essential during maintenance.
Every change either in the Scapolite source or the underlying infrastructure required for generating the artifacts for implementation and checking may lead to errors.
For example:




\begin{enumerate}
  \item Errors in the metadata introduced during maintenance may lead to rule omissions in the generated artifacts.
  \item Errors in the transformation from abstract to concrete machine-readable information may lead to faulty specifications, which in turn lead to faulty implementations and checks.
These transformation errors can originate from, e.g., bugs introduced during maintenance of the transformation library.
  \item Similarly, errors in the transformation to program code or other artifacts may lead to faulty implementations/checks.
\end{enumerate}

Further, we need to detect errors in a timely manner that are introduced by changes that have nothing to do with our process:

\begin{enumerate}
    \item Maintainers may misspecify the machine-readable information when making changes during maintenance.
    \item Changes in the target of hardening, e.g., upgrades of the OS, may invalidate or break a particular way of implementing or checking.
    \item Changes in execution environments for a created artifact, e.g., changes in a vulnerability scanner we generate a specification for, may invalidate the created artifact.
\end{enumerate}

Only a high automation degree allows us to run the required regression tests whenever a change occurs.

\section{Our approach to testing}
\label{sec:approach-to-testing}

\subsection{The Testing process}
\label{sec:test_process}

As pointed out in Section~\ref{sec:multitude_of_test_cases}, testing the implementation and checking of a security guide to a target is likely to require several test runs:
one for each combination of use-case, e.g., regular vs. high-security, used system, target-system revision, e.g., Windows release 1809 vs. 1909, and implementation or check runtime environments;
the latter either ingest some of the created artifacts, e.g., a test policy, or provide as external mechanisms a certain \textit{ground truth}.
We use, for example, the CIS-CAT scanner to verify implementations/checks generated for CIS baselines.
Nevertheless, we can also have different results for the same tools, e.g., because of different versions.

\subsubsection{Anatomy of typical test run}
\label{sec:test_run_anatomy}
A test run typically has the following shape:

\begin{description}[leftmargin=0.5em]
    \item[Run initial checks] Run checks on the unchanged system to establish the status quo \emph{before} the implementation.
    \item[Apply security settings] Execute the generated mechanism for implementing the desired security settings.
    \item[Carry out checks for compliance] Re-run checks against the changed system.
    \item[Revert settings] Revert the revertable settings to their initial status.
    \item[Check reverted settings] Check the status after we restored the settings' old state.
\end{description}

\subsubsection{Analysis of a test run}
\label{sec:testrun_analysis_tasks}

Relevant data that can be collected from such test runs are:

\begin{description}[leftmargin=0.5em]
    \item[Quantitative data] How many rules were successfully applied?
    For how many rules did the check return a success, a failure, a runtime problem, et cetera?
    \item[Detailed information] Which rules were successfully applied?
    For which rules was the check successful, a failure, ran into a problem, et cetera?
\end{description}

Analysis of the complete set of test runs for a specific
setting, i.e., a combination of use-case and target system,
usually entails two types of comparison:

\begin{description}[leftmargin=0.5em]
    \item[Comparisons within a test run] to find discrepancies, e.g.:
    \begin{itemize}[leftmargin=0.5em]
        \item A rule is reported as applied, but the check mechanism reports the rule as non-compliant.
        \item Two check mechanisms report different results for a rule.
        \item The check mechanism marked a rule as non-compliant before the implementation, compliant after the implementation, but still as compliant after the reverting.
    \end{itemize}
    \item[Comparison with previous test runs] to carry out regression tests: the newly collected data is compared with data from previous test executions.
    Were there changes?
    If so, are these \textit{desirable} changes, e.g., we improved an implementation or check that did not work before, or \textit{undesireable} changes, e.g., previously successful check does not succeed anymore.
\end{description}

\subsection{Our Approach to Test Automation}

In order to automate testing as much as possible, we implemented the following approach:
Our tooling automatically executes a machine-readable test specification on VMs created on-demand in AWS;
the tooling carries out the specified test activities, collects the raw data generated from implementation and check mechanisms, and automatically prepares summary data and data comparisons required to analyze the tests.

This complete automation of test activities allows an author or maintainer to carry out tests with no effort;
the extensive pre-processing of the test data enables them to see directly whether there are deviations from the expected results and enables them to focus on analyzing the cause of these deviations.

\subsubsection{Test Specification}
\label{sec:test_spec_overview}

With our YAML-based file format, we can define one or more test runs;
they are executed on different instances in parallel.
We specify:
\begin{itemize}[leftmargin=0.5em]
    \item for each test run, a sequence of activities such as implementing, checking, or reverting rules (cf. Section~\ref{sec:test_run_anatomy});
    \item for each activity, a list of so-called validations;
    each validation compiles data from the result or log files created by an activity (for example, validations can count successfully checked rules, collect these rules' identifiers, compare the current results to results of previous activities, et cetera);
    \item for each validation, the expected results (as basis for regression tests along with each validation)
\end{itemize}

The test specification file is kept under version control with the Scapolite sources for each security-configuration guide.

\subsubsection{Test Execution}
\label{sec:test_execution_overview}

We have implemented a test runner that is part of the DevOps pipeline that generates the artifacts for implementation and checks.
The test runner accesses the test specification file in the repository and executes the tests:

\begin{itemize}[leftmargin=0.5em]
    \item For each test run, the runner starts the required AWS image.
    \item The runner transfers the created artifacts and additional resources required for implementation/checking to the image.
    \item The runner uses Ansible to carry out the specified activities.
    \item In the end, the runner retrieves the created result/log files from each activity from the image, stops and destroys it.
\end{itemize}

\subsubsection{Preprocessing of test results}
\label{sec:test_preprocessing_overview}

As described in Section~\ref{sec:test_spec_overview}, we can specify validation tasks for each action carried out in the test run.
Hence, after the runner collected all raw data, the tooling carries out the validation tasks:
the required data is compiled, and a comparison to the expected results specified in the test specification file is carried out.

As a final step, our tooling commits (1) a detailed log, (2) a report of found deviations, (3) an updated test specification file with the current validation results, and (4) all raw data retrieved from the image to a staging repository.


\subsection{Test Specification}


\begin{listing}[!htb]
\begin{Verbatim}[commandchars=\\\{\}]
\PYG{n+nt}{os\PYGZus{}image}\PYG{p}{:} \PYG{l+lScalar+lScalarPlain}{Windows10}
\PYG{n+nt}{os\PYGZus{}image\PYGZus{}version}\PYG{p}{:} \PYG{l+lScalar+lScalarPlain}{1809}
\PYG{n+nt}{ciscat\PYGZus{}version}\PYG{p}{:} \PYG{l+lScalar+lScalarPlain}{v4.0.20}
\PYG{n+nt}{testruns}\PYG{p}{:}
\PYG{p+pIndicator}{\PYGZhy{}} \PYG{n+nt}{name}\PYG{p}{:} \PYG{l+lScalar+lScalarPlain}{1809 L2 High Security (...)}
\PYG{p+pIndicator}{\PYGZhy{}} \PYG{n+nt}{name}\PYG{p}{:} \PYG{l+lScalar+lScalarPlain}{1809\PYGZus{}Level1\PYGZus{}Corporate\PYGZus{}General\PYGZus{}use}
    \PYG{n+nt}{testrun\PYGZus{}ps\PYGZus{}profile}\PYG{p}{:} \PYG{l+lScalar+lScalarPlain}{L1\PYGZus{}Corp\PYGZus{}Env\PYGZus{}genUse}
    \PYG{n+nt}{testrun\PYGZus{}ciscat\PYGZus{}profile}\PYG{p}{:} \PYG{l+lScalar+lScalarPlain}{cisbenchmarks\PYGZus{}profile\PYGZus{}L1\PYGZus{}Corp\PYGZus{}Env\PYGZus{}genUse}
    \PYG{n+nt}{testrun\PYGZus{}benchmark\PYGZus{}filename}\PYG{p}{:} \PYG{l+lScalar+lScalarPlain}{CIS\PYGZus{}Win\PYGZus{}10\PYGZus{}1809\PYGZhy{}xccdf.xml}
    \PYG{n+nt}{activities}\PYG{p}{:}
    \PYG{p+pIndicator}{\PYGZhy{}} \PYG{n+nt}{id}\PYG{p}{:} \PYG{l+lScalar+lScalarPlain}{initial\PYGZus{}powershell\PYGZus{}check}
    \PYG{n+nt}{type}\PYG{p}{:} \PYG{l+lScalar+lScalarPlain}{ps\PYGZus{}scripts}
    \PYG{n+nt}{sub\PYGZus{}type}\PYG{p}{:} \PYG{l+lScalar+lScalarPlain}{check\PYGZus{}all}
    \PYG{n+nt}{validations}\PYG{p}{:}
    \PYG{p+pIndicator}{\PYGZhy{}} \PYG{n+nt}{sub\PYGZus{}type}\PYG{p}{:} \PYG{l+lScalar+lScalarPlain}{count}
        \PYG{n+nt}{expected}\PYG{p}{:}
        \PYG{n+nt}{blacklist\PYGZus{}rules}\PYG{p}{:} \PYG{l+lScalar+lScalarPlain}{0}
        \PYG{n+nt}{compliant\PYGZus{}checks}\PYG{p}{:} \PYG{l+lScalar+lScalarPlain}{75}
        \PYG{n+nt}{non\PYGZus{}compliant\PYGZus{}checks}\PYG{p}{:} \PYG{l+lScalar+lScalarPlain}{272}
        \PYG{n+nt}{empty\PYGZus{}checks}\PYG{p}{:} \PYG{l+lScalar+lScalarPlain}{2}
        \PYG{n+nt}{unknown\PYGZus{}checks}\PYG{p}{:} \PYG{l+lScalar+lScalarPlain}{2}
    \PYG{l+lScalar+lScalarPlain}{(...)}
    \PYG{p+pIndicator}{\PYGZhy{}} \PYG{n+nt}{id}\PYG{p}{:} \PYG{l+lScalar+lScalarPlain}{apply\PYGZus{}all}
    \PYG{n+nt}{type}\PYG{p}{:} \PYG{l+lScalar+lScalarPlain}{ps\PYGZus{}scripts}
    \PYG{n+nt}{sub\PYGZus{}type}\PYG{p}{:} \PYG{l+lScalar+lScalarPlain}{apply\PYGZus{}all}
    \PYG{n+nt}{blacklist\PYGZus{}rules}\PYG{p}{:} \PYG{p+pIndicator}{[}\PYG{n+nv}{R2\PYGZus{}2\PYGZus{}16}\PYG{p+pIndicator}{,} \PYG{n+nv}{R2\PYGZus{}3\PYGZus{}1\PYGZus{}1}\PYG{p+pIndicator}{,} \PYG{n+nv}{...}\PYG{p+pIndicator}{,} \PYG{n+nv}{R18\PYGZus{}9\PYGZus{}97\PYGZus{}2\PYGZus{}4}\PYG{p+pIndicator}{]}
    \PYG{n+nt}{validations}\PYG{p}{:}
    \PYG{p+pIndicator}{\PYGZhy{}} \PYG{n+nt}{sub\PYGZus{}type}\PYG{p}{:} \PYG{l+lScalar+lScalarPlain}{count}
        \PYG{n+nt}{expected}\PYG{p}{:}
        \PYG{n+nt}{applied\PYGZus{}automations}\PYG{p}{:} \PYG{l+lScalar+lScalarPlain}{336}
        \PYG{n+nt}{not\PYGZus{}applied\PYGZus{}automations}\PYG{p}{:} \PYG{l+lScalar+lScalarPlain}{4}
    \PYG{l+lScalar+lScalarPlain}{(...)}
    \PYG{p+pIndicator}{\PYGZhy{}} \PYG{n+nt}{id}\PYG{p}{:} \PYG{l+lScalar+lScalarPlain}{check\PYGZhy{}after\PYGZhy{}apply\PYGZhy{}all\PYGZhy{}with\PYGZhy{}ps}
    \PYG{n+nt}{type}\PYG{p}{:} \PYG{l+lScalar+lScalarPlain}{ps\PYGZus{}scripts}
    \PYG{n+nt}{sub\PYGZus{}type}\PYG{p}{:} \PYG{l+lScalar+lScalarPlain}{check\PYGZus{}all}
    \PYG{n+nt}{validations}\PYG{p}{:}
    \PYG{p+pIndicator}{\PYGZhy{}} \PYG{n+nt}{sub\PYGZus{}type}\PYG{p}{:} \PYG{l+lScalar+lScalarPlain}{by\PYGZus{}id}
        \PYG{n+nt}{result}\PYG{p}{:} \PYG{l+lScalar+lScalarPlain}{non\PYGZus{}compliant\PYGZus{}checks}
        \PYG{n+nt}{comment}\PYG{p}{:} \PYG{l+lScalar+lScalarPlain}{Correspond to blacklisted rules}
        \PYG{n+nt}{check\PYGZus{}ids}\PYG{p}{:} \PYG{p+pIndicator}{[}\PYG{n+nv}{R2\PYGZus{}2\PYGZus{}16}\PYG{p+pIndicator}{,} \PYG{n+nv}{R2\PYGZus{}3\PYGZus{}1\PYGZus{}1}\PYG{p+pIndicator}{,} \PYG{n+nv}{...}\PYG{p+pIndicator}{,} \PYG{n+nv}{R18\PYGZus{}9\PYGZus{}97\PYGZus{}2\PYGZus{}4}\PYG{p+pIndicator}{]}
    \PYG{l+lScalar+lScalarPlain}{(...)}
    \PYG{p+pIndicator}{\PYGZhy{}} \PYG{n+nt}{id}\PYG{p}{:} \PYG{l+lScalar+lScalarPlain}{check\PYGZus{}after\PYGZus{}apply\PYGZus{}all\PYGZus{}ciscat ...}
    \PYG{n+nt}{type}\PYG{p}{:} \PYG{l+lScalar+lScalarPlain}{ciscat}
    \PYG{n+nt}{validations}\PYG{p}{:}
    \PYG{p+pIndicator}{\PYGZhy{}} \PYG{n+nt}{sub\PYGZus{}type}\PYG{p}{:} \PYG{l+lScalar+lScalarPlain}{compare}
        \PYG{n+nt}{compare\PYGZus{}with}\PYG{p}{:} \PYG{l+lScalar+lScalarPlain}{check\PYGZhy{}after\PYGZhy{}apply\PYGZhy{}all\PYGZhy{}with\PYGZhy{}ps}
        \PYG{n+nt}{expected}\PYG{p}{:}
        \PYG{n+nt}{comment}\PYG{p}{:} \PYG{l+lScalar+lScalarPlain}{CISCAT error for 18.8.21.5}
        \PYG{n+nt}{rules\PYGZus{}failed\PYGZus{}only\PYGZus{}here}\PYG{p}{:} \PYG{p+pIndicator}{[}\PYG{n+nv}{R18\PYGZus{}8\PYGZus{}21\PYGZus{}5}\PYG{p+pIndicator}{,} \PYG{n+nv}{...}\PYG{p+pIndicator}{]}
        \PYG{n+nt}{rules\PYGZus{}unknown\PYGZus{}only\PYGZus{}here}\PYG{p}{:} \PYG{p+pIndicator}{[}\PYG{n+nv}{R1\PYGZus{}1\PYGZus{}5}\PYG{p+pIndicator}{,} \PYG{n+nv}{R1\PYGZus{}1\PYGZus{}6}\PYG{p+pIndicator}{,} \PYG{n+nv}{R2\PYGZus{}3\PYGZus{}10\PYGZus{}1}\PYG{p+pIndicator}{]}
        \PYG{n+nt}{rules\PYGZus{}unknown\PYGZus{}only\PYGZus{}there}\PYG{p}{:} \PYG{p+pIndicator}{[}\PYG{n+nv}{R18\PYGZus{}2\PYGZus{}1}\PYG{p+pIndicator}{,} \PYG{n+nv}{...}\PYG{p+pIndicator}{]}
        \PYG{n+nt}{rules\PYGZus{}passed\PYGZus{}only\PYGZus{}here}\PYG{p}{:} \PYG{p+pIndicator}{[]}
    \PYG{l+lScalar+lScalarPlain}{(...)}
\PYG{n+nt}{static}\PYG{p}{:}
\PYG{p+pIndicator}{\PYGZhy{}} \PYG{n+nt}{id}\PYG{p}{:} \PYG{l+lScalar+lScalarPlain}{validate\PYGZus{}json\PYGZus{}file}
    \PYG{n+nt}{type}\PYG{p}{:} \PYG{l+lScalar+lScalarPlain}{examine\PYGZus{}sfera\PYGZus{}automation\PYGZus{}json}
    \PYG{n+nt}{validations}\PYG{p}{:}
    \PYG{p+pIndicator}{\PYGZhy{}} \PYG{n+nt}{sub\PYGZus{}type}\PYG{p}{:} \PYG{l+lScalar+lScalarPlain}{count}
    \PYG{n+nt}{expected}\PYG{p}{:}
        \PYG{n+nt}{no\PYGZus{}automation}\PYG{p}{:} \PYG{l+lScalar+lScalarPlain}{1}
        \PYG{l+lScalar+lScalarPlain}{(...)}
    \PYG{p+pIndicator}{\PYGZhy{}} \PYG{n+nt}{sub\PYGZus{}type}\PYG{p}{:} \PYG{l+lScalar+lScalarPlain}{by\PYGZus{}id}
    \PYG{n+nt}{expected}\PYG{p}{:}
        \PYG{n+nt}{no\PYGZus{}automation}\PYG{p}{:} \PYG{p+pIndicator}{[}\PYG{n+nv}{R18\PYGZus{}2\PYGZus{}1}\PYG{p+pIndicator}{]}
        \PYG{n+nt}{same\PYGZus{}setting}\PYG{p}{:} \PYG{p+pIndicator}{[]}
\PYG{l+lScalar+lScalarPlain}{(...)}
\end{Verbatim}

\caption{A summarized version of a test specification file.}
\label{lst:testfile}
\end{listing}

\subsubsection{Structure of the test file}
\label{sec:testfile_structure}

Listing~\ref{lst:testfile} shows an exemplary test specification file.
As detailed in Section~\ref{sec:test_spec_overview}, each test run specifies several activities with a list of validations per activity
(colored lines are referred to below):

\begin{itemize}[leftmargin=0.5em]

    \item We specify two test runs (lines 5-6), one for the \textit{Level 2}, i.e., high-security, profile of a CIS Windows 10 (1809) Benchmark, the other one for the basic \textit{Level 1} profile.
    Here we only show parts of the latter.
    \item As explained in Section~\ref{sec:test_run_anatomy}, we start with a check of the unchanged system, using the generated PowerShell scripts (line 11).
    The first validation activity (lines 15--21) provides a count of the check result:
    how many rules were compliant, non-compliant, et cetera.
    Here, as in all the following examples, the values defined in the test specification file are the expected values taken from previous test runs.
    \item We continue using the generated PowerShell scripts to apply all rules (line 25) of the chosen \textit{Level 1} profile (line 7).
    As we will discuss in more detail in Section~\ref{sec:negative_effects}, we usually need to blacklist some rules (line 26) because there are rules breaking the test mechanism, e.g., by disrupting connections to the test machine.
    Again, amongst other things, we validate the number of successfully applied rules (line 30).
    \item We follow the rules' application with two check activities:
    we check with the generated PowerShell script (lines 33ff) and an external scanner provided by the CIS~\cite{ciscatpro} (lines 42ff).

    \begin{itemize}[leftmargin=0.5em]
        \item Here, we see an example of validating not just rule counts but the actual rule identifiers, e.g., as we examine the rules that our script reports as non-compliant (line 40).
        In line 39, a tester made a comment:
        the non-compliant rules correspond to the blacklisted rules (in line 26).
        \item We can also carry out other relevant comparisons automatically:
	    For example, in lines 45ff., the check results of the CIS scanner are compared with the results of our PowerShell script;
        in line 49, under the keyword \verb+rules_failed_only_here+, we see a list of rules which the CIS scanner reports as non-compliant, but our PowerShell scripts report as compliant.
        Again, a tester added a comment (line 48) about the reasons for the deviations.

        For example, for a specific rule, the CIS scanner requires that a particular setting should not be configured, even though the human-readable description of the rule requires that the setting should be disabled.
        Testers at \censor{Siemens} re-discovered systematic false positives like these repeatedly;
        by documenting such problems of external scanners, testers can better focus on actual deviations.
    \end{itemize}
    \item We also carry out static tests on the created artifacts (line 54ff.);
    the static tests are always carried out as the very first test activity.
    For example, we examine the created JSON file for entries without an automation (lines 60, 64) to catch errors during maintenance, leading to a failure when creating automations.
    Another valuable check is whether the same security setting is affected by several rules (line 65) since this often points to an error made during the rules' specification.
\end{itemize}

\subsubsection{Management of the test specification file}

When a test is carried out for the first time, the tester specifies the test runs, actions, and validations but leaves the fields about expected values empty since she does not know the expected values so far.
When the test is completed, the test infrastructure generates a version of the test specification file that contains all values from the tests' results.
The tester can use this version of the file as a basis for the following tests.

We manage the test specification file and the Scapolite sources that are the input to the pipeline in the same repository rather than at a separate location;
similar to a \verb+.gitlab-ci.yml+, we store the test specification file under \verb+.scapolite_tests.yml+.
Thus, during authoring/maintenance, when we create different branches, the test specification file is always part of the particular branch, drives the branch, and test results are fed back into the test specification file as expected results.

\subsection{Execution of Tests}

\subsubsection{Testing in the cloud}
\label{sec:testing_in_the_cloud}

Our test infrastructure started as a server equipped with VirtualBox\footnote{\url{https://www.virtualbox.org/}} for creating test images;
furthermore, we used Vagrant\footnote{\url{https://www.vagrantup.com/}} to manage image creation and destruction, Ansible for carrying out the test activities, and transferring data between the server and the images.

This approach, though well-suited for developing the test infrastructure, could not scale.
The combinatorial explosion in test cases that occurs for security-configuration guides often leads to many test cases.
Thus, we firstly must run all test runs for a single test in parallel to keep the time for executing a complete test acceptable.
Secondly, we need several authors/maintainers to work in parallel without the scarcity of test resources hindering them.

We, therefore, moved the testing process into the cloud and migrated from Oracle VirtualBox to AWS EC2\footnote{\url{https://aws.amazon.com/ec2/}}.
In the beginning, we had to overcome some initial problems caused by differences between VirtualBox and EC2 in credential management and the access of virtual machines.
Also, we had to redesign how we transfer data between the test runner and the images.
Using VirtualBox, the transfer of big files, e.g., the CIS-CAT scanner and a JVM to run it on, is essentially a local file-copy operation, whereas, with EC2, a naïve implementation would constantly transfer these files via the Internet from the local test runner to EC2.
We thus integrated an S3 bucket into our architecture, in which we host the files required for each test run:
hence, we transfer the data rather within the AWS data center than via the internet.

\subsubsection{Integration into DevOps pipeline}
\label{sec:devops_integration}

We generate the artifacts for implementing and checking security configurations from the Scapolite sources with a DevOps pipeline maintained as a GitLab-CI \emph{include file} within a dedicated repository.
For each Scapolite repository, we include this file into the GitLab-CI file;
because we factored out the actual code for the pipeline, we (1) keep the project's CI file very concise with only project-specific definitions, and (2) can carry out the maintenance of the pipeline via the single pipeline repository.

In code development, when changes are pushed to the code repository, tests are run changes are run.
In our case, however, each test entails the creation of several virtual machines, and the execution of a test run may take up to an hour.
We, therefore, chose to carry out only static tests for each push but require an active request by the author/maintainer for dynamic tests;
we realized this via a pipeline variable \verb+EXECUTE_TESTS+ passed to the pipeline.

\subsubsection{Dealing with negative effects of secure configurations on test execution}
\label{sec:negative_effects}

In Section~\ref{sec:testfile_structure}, we mentioned the \verb+blacklist+ definition required in test activities that implement security settings to preserve the  test infrastructure's functionality.
The infrastructure relies on specific mechanisms for accessing and manipulating the VM on which we carry out the tests.
Usually, the guides recommend disabling some of these mechanisms, e.g., firewall rules, rules restricting the use of stored credentials, et cetera, which may disrupt the WinRM functionality that Ansible uses.
If we implemented one of these rules, following test activities would cause Ansible to fail, with little or no information about why the activity failed.
In order to help the users with finding rules that break the test infrastructure, we implemented the following features:

\begin{itemize}[leftmargin=0.5em]
    \item Users can implement the rules in an \textit{apply} activity one by one rather than in bulk.
    A failure in execution can thus usually be attributed to the rule applied just before the failure occurred.
    \item To speed up test execution in this process of finding rules to blacklist, they can configure the rule implementation to start either at a specific rule or at the last rule contained in the blacklist; the guide specifies the rules' order.
    Unless a combination of rules causes an execution failure, this suffices to find all rules that must be blacklisted.
\end{itemize}

\subsection{User Feedback}

As shown above, we have highly automated the testing process itself.
However, the analysis of the test results still requires human interaction.
It is thus necessary to present the test results such that they provide the user with a concise overview of whether something went wrong and allow easy access to the raw data necessary for an in-depth analysis of problems uncovered by the test.

\subsubsection{Summary Report}

Once we executed all test runs and the analyses and comparisons specified for each activity have been carried out, our tooling generates a summary report providing concise information for each activity:
\begin{enumerate}
    \item Did failures occur during an activity, e.g., because a setting interrupted the connection to the virtual image and the activity could not be completed?
    \item If no failure occurred, did the test yield the expected results as documented in the test specification file?
    \item Where possible:
    if the test yielded different results, did the test show an \textit{improvement}?
    Were more rules implemented successfully than during the previous test run?
\end{enumerate}

With item 3) we intend to provide the user with an initial assessment of the test results.
This, however, requires a definition of what constitutes an improvement/degradation.
The users can specify in the test specification file what an improvement should be along with the expected data.
For example, the key-value pair \verb+improvement:rise+ in combination with a validation that counts results, e.g., the number of successfully implemented rules or of checks showing compliance, signifies that a reported higher number constitutes an improvement;
\verb+improvement:fall+ would do the opposite.
If no condition for an improvement is specified, a degradation is reported by default if the test results do not match the expected data.


\subsubsection{Documentation of full results}

In case a deeper analysis of the results becomes necessary, the users can access detailed information about found deviations for each validation step:
Listing~\ref{lst:example_difference_report} provides an example of how a deviation is reported.
Furthermore, users can access the raw data for each activity within a staging repository containing the generated artifacts.
Thus, all relevant data are provided at one location.
Also, they can use different mechanisms provided by \emph{git} and \emph{GitLab} such as viewing differences between test executions, e.g., within the generated artifacts, during the analysis of the test results.

\begin{listing}[!htb]
\begin{Verbatim}[commandchars=\\\{\}]
CRITICAL \PYGZhy{} Validation failed, SAME numbers, but DIFFERENT IDs (IMPROVEMENT: \PYGZsq{}fall\PYGZsq{})!
    Expected and confirmed(found) \PYGZsq{}unknown\PYGZus{}checks\PYGZsq{} IDs: \PYGZob{}\PYGZsq{}R18\PYGZus{}2\PYGZus{}1\PYGZsq{}, \PYGZsq{}R2\PYGZus{}3\PYGZus{}1\PYGZus{}6\PYGZsq{}, \PYGZsq{}R2\PYGZus{}2\PYGZus{}21\PYGZsq{}, \PYGZsq{}R2\PYGZus{}3\PYGZus{}1\PYGZus{}5\PYGZsq{}\PYGZcb{}
    Expected \PYGZsq{}unknown\PYGZus{}checks\PYGZsq{} IDs, but not found: \PYGZob{}\PYGZsq{}R2\PYGZus{}3\PYGZus{}11\PYGZus{}3\PYGZsq{}\PYGZcb{}
    Found \PYGZsq{}unknown\PYGZus{}checks\PYGZsq{} IDs, but not expected: \PYGZob{}\PYGZsq{}R19\PYGZus{}7\PYGZus{}41\PYGZus{}1\PYGZsq{}\PYGZcb{}
\end{Verbatim}

    \caption{Example report of a difference between test results and expected results.}
    \label{lst:example_difference_report}
\end{listing}

\subsubsection{Further automation}

We provide further support to the users if they need to re-test several guides, e.g., when the transformation mechanism was updated.
These command-line scripts that use the \emph{GitLab} API include tasks like:

\begin{itemize}[leftmargin=0.5em]
    \item starting pipelines in parallel for several guides;
    \item informing about the pipelines' status;
    \item compiling an overview with the results of all test pipelines;
    \item showing differences between the newly-generated artifacts and the latest published version for each guide;
\end{itemize}

By automating repetitive manual tasks carried out for each guide, we achieve that tests are executed frequently.
Especially small or seemingly \textit{harmless} changes are now more often tested because we lowered the effort for starting the tests and analyzing the test results for more than one guide significantly.

\section{Related Work}
\label{sec:relatedwork}

First, we present the current work on configuration management in general and security hardening in particular.
Second, we discuss approaches similar to our testing approach.

In past, reseachers investigated heavily in misconfiguration in general, and security misconfiguration in particular ~\cite{Dietrich:2018:ISO:3243734.3243794,Tang:2015:HCM:2815400.2815401,Continella:2018:THB:3274694.3274736,Jha2017,Xu:2015:SAT:2775083.2791577}.
Dietrich et al.~~\cite{Dietrich:2018:ISO:3243734.3243794} show in their study that security misconfigurations are very common and a severe problem.
According to their data, manual configuration, vague or no process, and poor internal documentation are the main environmental factors that we could solve with a better approach and tooling.

Many researchers investigated  how we can detect and remove misconfigurations~\cite{Rahman:2019:SSS:3339505.3339528,Santolucito:2017:SCF:3152284.3133888,Keller2008,Su:2007:AIC:1323293.1294284}.
Rahman et al.~\cite{Rahman:2019:SSS:3339505.3339528} analyzed thousands of IaC scripts to identify insecure configurations;
the framework \textit{ConfigV}~\cite{Santolucito:2017:SCF:3152284.3133888} learns good configuration settings based on given configuration files. 
Depending on the guide's target, such techniques could be used to develop the guide or check for problems with the chosen configuration settings by applying them to the generated implementation artifacts.
SPEX~\cite{Xu:2013:BUM:2517349.2522727} examines the source code of programs in order to find security-related configurations and would thus be useful in the creation of public guides for open-source software as well as internal guides for one's products. 

The creation of automated implementation/check mechanisms becomes much easier when a unified framework for setting and checking configurations for a software product is in place.
The Elektra framework~\cite{Raab2020}, for example, unifies how we can access configuration settings and creates a central structure for accessing and manipulated configuration settings.
Xu et al.~\cite{Xu2016} developed a similar approach to Elektra.
Furthermore, they showed~\cite{Xu:2015:HYG:2786805.2786852} convincingly that the configuration's complexity is overwhelming users and systems administrators.
The results of the study underline how important security experts and security guides are in supporting the administrators.

The ComplianceAsCode project~\cite{complianceAsCode,preisler:haicman:UsenixLisa:2017} is very close to the presented approach.
The authors maintain their security-configuration guides for various Linux systems in a git repository and represent every rule with one file.
This file references other files, e.g., with scripts for automated checking.
Nevertheless, some drawbacks prevented us from using ComplianceAsCode.
First, their focus on Linux-based operating systems did not support our initial, primary use of Windows hardening.
Second, in contrast to ComplianceAsCode, we try to generate as much as possible from a single abstract specification, whereas ComplianceAsCode maintains a check in OVAL and the implementation mechanism(s) for each setting in a different language.
Nonetheless, it would ease the security configuration enormously if the publishers distributed their guides in a format akin to ComplianceAsCode so that documentation, check \emph{and} implementation are more aligned.

Software testing is a well-researched discipline, and every year, new articles add more information to the general knowledge~\cite{9286051,9286080,9159077,9159091,9159060,9159084,9286071,9286121,9285999,9286104}.
Therefore, we can only refer to a fraction of all available and valuable testing research.
Many researchers, e.g., \cite{9159084,10.1145/3395363.3397365} use sophisticated testing approaches to find security-relevant bugs or leaks in software.
In contrast, we use testing approaches to find bugs in the security-configuration guides, not the software itself.
In industry, there is a strong need for automated testing, especially in the DevOps scenarios~\cite{9155725}.
Also, there are some obstacles to overcome, e.g., when testing a software's graphical user interface~\cite{9155991}.
Since we use our approach productively at \censor{Siemens}, we had to overcome similar problems as the researchers above.
The closest research to our process of testing security-configuration guides is the work of Spichkova et al.~\cite{SPICHKOVA20203610}.
Their tool VM2 creates VM images and hardens them automatically with given security-configuration guides.
They also use the CIS's guides, but they see guides as given and immutable, whereas we include in our approach the constant update and maintenance of the guides to adjust them to a company's security policy.
Furthermore, they focus on the combination of Linux-based OSs and Ansible.
In contrast, the diversity of technologies within \censor{Siemens} forced us to support different application and check modes in our approach.

\section{Conclusion}
\label{sec:conclusion}

We have developed an approach towards authoring and maintaining machine-readable security-configuration guides that allows us to extend the DevOps principle of Continuous Integration to this domain.
We achieved this by creating the Scapolite format that enables authors to combine human-readable information with machine-readable information on security-configuration requirements.
The latter then serve as input for a process that (1) automatically generates artifacts for implementation and checking and (2) tests the created artifacts. 
Because the authors can specify the rules on an abstract level and thus do not have to manage such artifacts in parallel, we could significantly reduce the risk of errors because of manual errors and inconsistencies.

Due to the high degree of automation in our proposed process, we test the security-configuration guides and their generated artifacts much more frequently during authoring and maintenance than in the normal case.
As a result, we detect the majority of problems before the release of a security-configuration guide.

In summary, our approach to security hardening via machine-readable security-configuration guides combined with the automated testing allows us to publish automated, well-tested mechanisms for implementing and checking along with the guide. 
Consequently, compliance with these configurations can be reached in a more timely and less error-prone manner, leading to better-secured systems. 




\bibliographystyle{ACM-Reference-Format}
\bibliography{main}

\end{document}